\documentstyle[12pt]{article}

\newcommand {\e} {\mbox{\rm e}}

\newcommand {\nn}    {\nonumber}
\newcommand {\vs}[1]  { \vspace*{#1 cm} }

\newcounter{eq}
\newcounter{sc}

\newcommand {\MPL}  {Mod.Phys.Lett.}
\newcommand {\NP}   {Nucl.Phys.}
\newcommand {\PL}   {Phys.Lett.}
\newcommand {\PR}   {Phys.Rev.}
\newcommand {\PRL}   {Phys.Rev.Lett.}

\newcommand {\AP}   {Ann.of Phys.}

\def\overleftrightarrow#1{\vbox{\ialign{##\crcr
 $\leftrightarrow$\crcr\noalign{\kern-1pt\nointerlineskip}
 $\hfil\displaystyle{#1}\hfil$\crcr}}}

\newlength{\minitwocolumn}
\begin{document}

\begin{flushright}
EDO-EP-38\\
March, 2001\\
\end{flushright}
\vspace{30pt}

\pagestyle{empty}
\baselineskip15pt

\begin{center}
{\large\bf A New Mechanism for Trapping of Photon

 \vskip 1mm
}

\vspace{20mm}

Ichiro Oda
          \footnote{
          E-mail address:\ ioda@edogawa-u.ac.jp
                  }
\\
\vspace{10mm}
          Edogawa University,
          474 Komaki, Nagareyama City, Chiba 270-0198, JAPAN \\

\end{center}


\vspace{15mm}
\begin{abstract}
We propose a new mechanism for trapping bulk gauge field, giving rise
to a massless photon on a flat Minkowski 3-brane in the Randall-Sundrum 
model in five space-time dimensions. The mechanism we propose employs the 
topological Higgs mechansim where a topological term and a 3-form gauge
potential play an important role.
This new mechanism might be considered as a gauge field's analog of 
the localization of bulk fermions with the mass term of a 'kink' profile.

\vspace{15mm}

\end{abstract}

\newpage
\pagestyle{plain}
\pagenumbering{arabic}

\rm

The gravity-localized models in a brane world offer an opportunity
for a new solution to the hierarcy problem, an alternative 
compactification scenario and the cosmological constant problem 
and so on \cite{Randall1, Randall2}.
The crucial ingredient in the models is the localization mechanism
for all the familiar matter and gauge fields in addition to the
graviton on a brane. 
In superstring theory, matter and gauge fields are naturally confined
to D3 branes due to open strings ending on the branes while the gravity
is free to propagate in a bulk space-time due to closed strings
living in the bulk \cite{Pol}.
On the other hand, in local field theory, it is known that in contrast
with the other fields, the gauge fields cannot be localized on a brane
by a gravitational interaction \cite{Dvali, Pomarol, Rizzo, Bajc, Oda0}. 
(For a review see \cite{Oda1}.)
The non-localization of the bulk gauge fields on a brane is one of 
fatal drawbacks associated with the gravity-localized models since 
there certainly exists a massless '$\it{photon}$' in our world.

The aim of the present paper is to propose a new localization mechanism
of the gauge fields on a 3-brane from the viewpoint of local field
theory in the framework of the Randall-Sundrum model \cite{Randall2}. 
The possibility of constructing such a new mechanism has stemmed from an 
attempt of making a gauge field's analog of the localization mechanism
of the bulk fermions. So let us explain briefly our background 
idea behind the new mechanism in connection with the fermion localization. 

The problem of localizing fermion zero-modes on a domain wall has been 
already solved by Jackiw and Rebbi \cite{Jackiw} where the Yukawa interaction 
$\bar{\Psi} \Psi \Phi$ interpolating two different vacua at each side of 
a domain wall yields the mass term with a 'kink' profile
$m \varepsilon(r) \bar{\Psi} \Psi$ with the definitions of $r$ being a 
transverse dimension to the domain wall and $\varepsilon(r)$ being the 
step function given by $\varepsilon(r) = \frac{r}{|r|}$ and 
$\varepsilon(0) = 0$. 
Because of this type of mass term, the bulk fermions can be localized on 
the domain wall by a gravitational interaction. Then it is natural to ask 
ourselves whether we can apply this localization mechanism of fermions to 
gauge fields or not. However, it turns out that the introduction of a mass 
term in a simple manner does not lead to any localization of the bulk gauge 
fields \cite{Pomarol}.

To pursue an analogy to some extent, let us write down the actions for both 
spin-1/2 spinor and spin-1 vector fields. 
After taking a vacuum expectation value of a 'kink' profile,
the action for the spin-1/2 spinor field reads
\begin{eqnarray}
S = - \int d^5 x \sqrt{-g} \bar{\Psi} i ( \Gamma^M D_M + m
\varepsilon(r)) \Psi,
\label{1}
\end{eqnarray}
where the covariant derivative is defined as $D_M \Psi = (\partial_M
+ \frac{1}{4} \omega_M^{AB} \gamma_{AB}) \Psi$ with $\omega_M^{AB}$
being the spin connections.
On the other hand, after spontaneous symmetry breakdown, the relevant
action for the spin-1 $U(1)$ vector field is given by
\begin{eqnarray}
S = \int d^5 x \sqrt{-g} ( -\frac{1}{4} g^{M N} g^{R S} 
F_{MR} F_{NS} + \frac{1}{2} m^2 g^{M N} A_M A_N), 
\label{2}
\end{eqnarray}
where $F_{MN} = \partial_M A_N - \partial_N A_M$. As mentioned above,
in the Randall-Sundrum model \cite{Randall1, Randall2} the zero-mode
of the spinor field is localized on a brane, whereas the one of the 
gauge field is not so. 
Then one might wonder what differences there are between the two
actions.
One obvious difference lies in the form of mass term where the mass term
takes a linear form depending on the step function in (\ref{1}) 
while it is a quadratic form in (\ref{2}). The other important difference
is that the action (\ref{1}) is invariant under gauge transformation 
while the action (\ref{2}) is implicitly broken owing to spontaneous
symmetry breakdown. 

At this stage, it is of interest to ask ourselves if there is an 
action for the gauge field which is not only manifestly invariant 
under the gauge transformation but also has a linear mass term. 
To the best of our knowledge,
there is only one action, which is nothing but the pure gauge theory 
with a topological term. Recall that in this theory the mass of the
gauge field is generated by the so-called topological Higgs mechanism
owing to the existence of a topological term instead of the conventional
Higgs mechanism \cite{Schonfeld, Deser, Oda2}.
Therefore, in the present paper, we shall consider this theory 
and study the localization of the zero-mode of the bulk gauge field 
on a brane in the framework of the Randall-Sundrum model. 
Surprisingly enough,
we will see that the theory provides us an ingenious mechanism for
the localization of the gauge field on a flat 3-brane.

We shall start by fixing the model setup.  
The metric ansatz we adopt is of the Randall-Sundrum form \cite{Randall2}:
\begin{eqnarray}
ds^2 &=& g_{MN} dx^M dx^N  \nn\\
&=& \e^{-A(r)} \eta_{\mu\nu} dx^\mu dx^\nu + dr^2,
\label{3}
\end{eqnarray}
where $M, N, ...$ denote five-dimensional space-time indices and 
$\mu, \nu$, ...four-dimensional brane ones. The metric on the brane
$\eta_{\mu\nu}$ denotes the four-dimensional flat Minkowski metric with
signature $(-,+,+,+)$. Moreover, $A(r) = 2 k |r|$ where $k$ is a positive 
constant and the fifth dimension $r$ runs from $-\infty$ to $\infty$.
We have the physical situation in mind where a single flat 3-brane sits
at the origin of the fifth dimension, $r = 0$, and then ask if the
bulk $U(1)$ gauge field can be localized on the brane by a gravitational
interaction. If we find a normalizable zero-mode of a bulk field with 
exponential falloff, we regard the zero-mode as a local field on our
world corresponding to the bulk field. In this article, we will assume
that the background metric is not modified by the presence of the bulk 
fields, that is, we will neglect the back-reaction on the metric from
the bulk fields.

Now we are ready to consider the Maxwell's action for $U(1)$ massless
gauge field plus the actions for a 3-form potential and a topological
term whose total action is explicitly given by 
\begin{eqnarray}
S &=& \int d^5 x [ -\frac{1}{4} \sqrt{-g} g^{M_1 N_1} g^{M_2 N_2} 
F_{M_1 M_2} F_{N_1 N_2} - \frac{1}{48} \sqrt{-g} g^{M_1 N_1} g^{M_2 N_2} 
g^{M_3 N_3} g^{M_4 N_4} \times \nn\\
&{}& H_{M_1 M_2 M_3 M_4} H_{N_1 N_2 N_3 N_4}
+ \frac{m}{6} \varepsilon^{M_1 M_2 M_3 M_4 M_5} C_{M_1 M_2 M_3}
\partial_{M_4} A_{M_5} ], 
\label{4}
\end{eqnarray}
where $F_{MNPQ} = 4 \partial_{[M} C_{NPQ]} = \partial_M C_{NPQ} - 
\partial_N C_{MPQ} + \partial_P C_{MNQ} - \partial_Q C_{MNP}$,
and $m$ is a positive constant.
This action has the following gauge symmtries as well as reducible
symmetries \cite{Batalin}:
\begin{eqnarray}
\delta A_M &=& \partial_M \lambda, \nn\\
\delta C_{MNP} &=& \partial_{[M} \varepsilon_{NP]}, \nn\\
\delta \varepsilon_{MN} &=& \partial_{[M} \varepsilon_{N]}, \nn\\
\delta \varepsilon_M &=& \partial_M \varepsilon.
\label{5}
\end{eqnarray}
Note that the gauge symmetry with respect to $A_M$ prohibits
the mass term of a 'kink' profile $m \varepsilon(r)$ in the action.

Then, the equations of motion read 
\begin{eqnarray}
\partial_{M_2} (\sqrt{-g} g^{M_1 N_1} g^{M_2 N_2} F_{N_1 N_2})
- \frac{m}{6} \varepsilon^{M_1 M_2 M_3 M_4 M_5} \partial_{M_2}
C_{M_3 M_4 M_5} &=& 0, \nn\\
\partial_{M_4} (\sqrt{-g} g^{M_1 N_1} g^{M_2 N_2} g^{M_3 N_3} 
g^{M_4 N_4} H_{N_1 N_2 N_3 N_4}) - m \varepsilon^{M_1 M_2 M_3 M_4 M_5} 
\partial_{M_4} A_{M_5} &=& 0. 
\label{6}
\end{eqnarray}
As the gauge conditions of the symmetries (\ref{5}), we shall take 
\begin{eqnarray}
A_r(x^M) &=& 0, \nn\\
C_{MNr}(x^M) &=& 0.
\label{7}
\end{eqnarray}
In particuar, note that the number of the latter gauge conditions precisely 
coincides with that of symmtries, which is $\frac{5 \times 4}{2} - 5 + 1
= \frac{4 \times 3}{2} = 6$. With these gauge conditions, we wish to 
look for a zero-mode solution with the forms of
\begin{eqnarray}
A_\mu(x^M) &=& a_\mu(x^\lambda) u(r), \nn\\
C_{\mu\nu\rho}(x^M) &=& c_{\mu\nu\rho} (x^\lambda) u(r).
\label{8}
\end{eqnarray}
where we assume the equations of motion in four-dimensional flat
space-time:
\begin{eqnarray}
\partial^\mu f_{\mu\nu} = \partial^\mu h_{\mu\nu\rho\sigma} = 0,
\label{9}
\end{eqnarray}
with the definitions of $f_{\mu\nu} = 2 \partial_{[\mu} a_{\nu]}$
and $h_{\mu\nu\rho\sigma} = 4 \partial_{[\mu} c_{\nu\rho\sigma]}$.

With these ansatzs, the equations of motion (\ref{6}) reduce to 
the following differential equations:
\begin{eqnarray}
a^\mu \partial_r ( \e^{- A(r)} \partial_r u(r) )
- \frac{m}{6} \varepsilon^{\mu\nu\rho\sigma} c_{\nu\rho\sigma}
\partial_r u(r) = 0,
\label{10}
\end{eqnarray}
\begin{eqnarray}
\e^{- A(r)} \partial^\mu a_\mu \partial_r u(r)
- \frac{m}{6} \varepsilon^{\mu\nu\rho\sigma} 
\partial_\mu c_{\nu\rho\sigma} u(r) = 0,
\label{11}
\end{eqnarray}
\begin{eqnarray}
c^{\mu\nu\rho} \partial_r ( \e^{A(r)} \partial_r u(r) )
- m \varepsilon^{\mu\nu\rho\sigma} a_\sigma \partial_r u(r) = 0,
\label{12}
\end{eqnarray}
\begin{eqnarray}
\e^{A(r)} \partial_\rho c^{\mu\nu\rho} \partial_r u(r)
- m \varepsilon^{\mu\nu\rho\sigma} \partial_\rho a_\sigma u(r) = 0.
\label{13}
\end{eqnarray}
Since Eqs. (\ref{11}), (\ref{13}) are the first-order differential
equations with respect to $x^\mu$-differentiation, we regard them
as the gauge conditions in four-dimensional space-time. (Later, we
will comment on these equations.) Then, the equations which we have 
to solve are Eqs. (\ref{10}), (\ref{12}). From the two equations, 
we can derive a differential equation to $u(r)$:
\begin{eqnarray}
(\partial_r u(r))^2 - m^2 u^2(r) = 0.
\label{14}
\end{eqnarray}
Given our physical setup that a single brane sits at $r = 0$ where
there is a $\delta$-functional source, 
a general solution to (\ref{14}) is given by
\begin{eqnarray}
u(r) = c \e^{\pm m \varepsilon(r) r},
\label{15}
\end{eqnarray}
where $c$ is an integration constant. In this respect,
it is worthwhile to notice that we have effectively selected
the mass term with a 'kink' profile as a solution to the equations of
motion. When we impose the boundary conditions such that
$u(\pm \infty) = 0, u(0) = u_0$, a special solution takes the form
\begin{eqnarray}
u(r) = u_0 \e^{- m \varepsilon(r) r}.
\label{16}
\end{eqnarray}
 
We are now willing to check that this zero-mode solution of the gauge
field as well as the 3-form potential leads to a normalizable mode
and the localization on a brane. For this, let us plug Eqs. (\ref{7}),
(\ref{8}) into the starting action (\ref{4}), and then a simple
calculation yields
\begin{eqnarray}
S^{(0)} &=& \int d^4 x \int_{-\infty}^{\infty} dr
[ -\frac{1}{4} f^2_{\mu\nu} u^2(r) - \frac{1}{2} a_\mu a^\mu 
\e^{-A} (\partial_r u)^2 \nn\\
&-& \frac{1}{48} \e^{2A} h^2_{\mu\nu\rho\sigma} u^2(r) 
- \frac{1}{12} \e^{A} c^2_{\mu\nu\rho} (\partial_r u)^2], 
\label{17}
\end{eqnarray}
where the topological term has disappeared from the above action 
by integration over $r$. Whether the zero-mode (\ref{16}) is normalizable
or not can be checked by evaluating each $r$-integral in (\ref{17}).
In particular, the first $r$-integral in front of the kinetic term
of the gauge field reads
\begin{eqnarray}
I_1 \equiv \int_{-\infty}^{\infty} dr u^2(r) = \frac{u_0^2}{m},
\label{18}
\end{eqnarray}
where $m > 0$ is utilized. The finiteness of this integral means
that the zero-mode (\ref{16}) is certainly a normalizable mode at
least for the gauge field. The second $r$-integral is similarly 
evaluated as
\begin{eqnarray}
I_2 \equiv \int_{-\infty}^{\infty} dr \e^{-A} (\partial_r u)^2
= \frac{m u_0^2}{m + k},
\label{19}
\end{eqnarray}
where $m > 0$ and  $k >0$ are used. The third $r$-integral becomes
\begin{eqnarray}
I_3 \equiv \int_{-\infty}^{\infty} dr \e^{2A} u^2(r)
= \frac{m u_0^2}{m - 2k}.
\label{20}
\end{eqnarray}
Note that this integral takes the above finite value if $m - 2k > 0$,
whereas it diverges if $m - 2k \le 0$. The final integral is given by
\begin{eqnarray}
I_4 \equiv \int_{-\infty}^{\infty} dr \e^A (\partial_r u)^2
= \frac{m u_0^2}{m - k}.
\label{21}
\end{eqnarray}
This integral also takes the finite value if $m - k > 0$, but it
becomes divergent if $m - k \le 0$.

First of all, let us consider the case that all the $r$-integrals 
have the finite values, for which the inequality $m - 2 k > 0$ must 
be satisfied.
Then in order to transform the kinetic terms to a canonical form, 
let us redefine the fields as
\begin{eqnarray}
\frac{u_0}{\sqrt{m}} a_\mu &\rightarrow& a_\mu, \nn\\
\frac{u_0}{\sqrt{m - 2k}} c_{\mu\nu\rho} &\rightarrow& 
c_{\mu\nu\rho}.
\label{22}
\end{eqnarray}
As a result, the action (\ref{17}) reduces to
\begin{eqnarray}
S^{(0)} = \int d^4 x 
[ -\frac{1}{4} f^2_{\mu\nu} - \frac{1}{2} \frac{m^3}{m + k} a_\mu a^\mu 
- \frac{1}{48} h^2_{\mu\nu\rho\sigma}  
- \frac{1}{12} \frac{m^2 (m - 2k)}{m - k} c^2_{\mu\nu\rho} ]. 
\label{23}
\end{eqnarray}
Moreover, the normalized zero mode of the gauge field in a flat 
space-time is of the form
\begin{eqnarray}
\hat{u}(r) = \frac{1}{\sqrt{I_1}} u(r) = \sqrt{m} 
\e^{- m \varepsilon(r) r}.
\label{24}
\end{eqnarray}
Here we encounter a problem. Recall that there is now the condition
$m - 2 k > 0$ where both $m$ and $k$ are positive constants. 
Together with it, the massless condition of '$\it{photon}$' gives us
the condition $m \approx 0$. Then the normalized zero mode of the 
gauge field, $\hat{u}(r)$ in (\ref{24}), spreads more widely in a bulk,
in other words, the brane gauge field is not sharply localized near
a brane. This situation is very similar to the case of the locally
localized model \cite{Oda3, Oda4, Oda5} where 'small extra dimensions' 
scenario was proposed in order to avoid this problem.
   
Is there any possibility to circumvent this 'small extra dimensions' 
scenario? Interestingly enough, in the case at hand, there $\it{is}$
an alternative possibility of trapping the bulk gauge field on a brane
without invoking the 'small extra dimensions' scenario. To do so,
let us notice that the root of the problem exists in the inequality
$m - 2 k > 0$ which has stemmed from the condition that the zero-mode
of a 3-form potential should be normalizable. We can now relax this 
condition if we consider the situation where the 3-form potential
does not have a normalizable zero-mode, thereby implying that the
3-form potential is not localized near a brane but resides in a whole bulk.
In this situation, we have a condition $m - 2 k \le 0$ and the 
brane action is effectively given by
\begin{eqnarray}
S^{(0)} = \int d^4 x 
[ -\frac{1}{4} f^2_{\mu\nu} - \frac{1}{2} \frac{m^3}{m + k} a_\mu a^\mu ]. 
\label{25}
\end{eqnarray}
Then the massless condition of the gauge field requires the
relation $\frac{m^3}{m + k} \ll 1$. These conditions are simply
satisfied by taking $k \gg m$. Hence, in this sense,
we have succeeded in getting a massless '$\it{photon}$' which
is sharply localized on a flat 3-brane.
Note that under the condition $k \gg m$, the 3-form potential resides 
in a bulk away from a brane since the zero-mode in a flat space has 
the behavior of $v(r) \sim \e^{-(m - 2k) \varepsilon(r) r}$.

Now some remarks are in order. The first remark is related to the
gauge conditions (\ref{11}) and (\ref{13}) in four dimensions.
Owing to the solution (\ref{16}), these gauge conditions must take
the different forms at each side of a 3-brane. Incidentally, we can 
rewrite them as
\begin{eqnarray}
\partial^\mu a_\mu - \frac{m}{6} \varepsilon(r) \e^{2k|r|}
\varepsilon^{\mu\nu\rho\sigma} \partial_\mu c_{\nu\rho\sigma} &=& 0,
\nn\\
\partial_\rho c^{\mu\nu\rho} - \varepsilon(r) \e^{-2k|r|} 
\varepsilon^{\mu\nu\rho\sigma} \partial_\rho a_\sigma &=& 0.
\label{26}
\end{eqnarray}
Thus, on the brane, they reduce to the usual gauge conditions
$\partial^\mu a_\mu = \partial_\rho c^{\mu\nu\rho} = 0$.

The second remark refers to the consistency of the equations of motion
in four dimensions. In deriving the differential equations 
(\ref{10})-(\ref{13}), we have assumed the four-dimensional equations
(\ref{9}). For self-consistency, we have to check that these equations
are indeed valid in four dimensions. It is clear that the equations of
motion, $\partial^\mu f_{\mu\nu} = 0$ hold as far as $\frac{m^3}{m + k} 
\ll 1$ as seen in (\ref{25}). On the other hand, the equations of motion,
$\partial^\mu h_{\mu\nu\rho\sigma} = 0$ need some attention since 
$c_{\mu\nu\rho}$ now reside in the region away from a brane. For this,
we shall consider the action (\ref{17}), from which we have the equations
of motion for $c_{\mu\nu\rho}$:
\begin{eqnarray}
\partial^\mu h_{\mu\nu\rho\sigma} - \e^{-2k|r|} m^2
c_{\nu\rho\sigma} = 0.
\label{27}
\end{eqnarray}
Since $c_{\mu\nu\rho}$ live in the region away from a brane where
$r \gg 1$ and we have the condition $k \gg 1$, 
Eq. (\ref{27}) means the desired equations, 
$\partial^\mu h_{\mu\nu\rho\sigma} = 0$.

As a final remark, it is worthwhile to point out that the starting
action (\ref{4}) often appears in the context of superstring
theory. For instance, it is well known that we need the term
$\int_{M_{10}} C \wedge X_8$ in the string-theory effective
action for the Green-Schwarz anomaly cancellation \cite{Green}.
This complicated term leads to a topological term $m \int_{M_5} C_3 
\wedge F$ upon compactification to five dimensions. Also, in
type IIA superstring theory, there is the gauge field together with 
a 3-form potential from the Ramond-Ramond sector with a topological
term \cite{Pol}. Hence, the new localization mechanism for 
'$\it{photon}$' which we have proposed in this paper may have
some applications within superstring theory.

In conclusion, we have proposed a new localization mechanism for
the gauge fields in the Randall-Sundrum gravity-localized model.
This new mechanism is very similar to that of fermions in the sense
that in the both mechanisms the zero-modes share the same form and 
the presence of the mass term of a 'kink' profile plays an essential 
role for trapping the zero-modes of the bulk fields on a flat 
Minkowski brane.
So we have called the present localization mechanism a gauge field's
analog of the localization of fermions in the abstract. 
The model at hand naturally appears in superstring theory,
so the present localization mechanism might also shed new light on 
superstring theory in addition to local field theory.
In near future, we wish to understand various aspects of this
new mechanism in connection with superstring theory.

\vs 1


\end{document}